\documentclass[aps,pra,twocolumn,superscriptaddress,showpacs,floatfix]{revtex4-1}
\usepackage{graphicx,dcolumn,bm,hyperref,amsmath,amssymb,xspace,epsfig,float,array,multirow,amsfonts,color}
\usepackage[caption=false]{subfig}
\usepackage[vcentermath]{youngtab}
\usepackage{braket}
\usepackage[normalem]{ulem}
\usepackage{ytableau}
\newcommand{\Id}{{\mbox{l\hspace{-0.52em}1}}}

\begin{document}

\title{A universal graph description for one-dimensional exchange models}

\author{Jean Decamp}
\affiliation{MajuLab, CNRS-UCA-SU-NUS-NTU International Joint Research Unit, Singapore}
\affiliation{Centre for Quantum Technologies, National University of Singapore, 117543 Singapore, Singapore}
\affiliation{Department of Physics, National University of Singapore, Singapore 117542, Singapore}

\author{Jiangbin Gong}
\affiliation{Department of Physics, National University of Singapore, Singapore 117542, Singapore}

\author{Huanqian Loh}
\affiliation{Centre for Quantum Technologies, National University of Singapore, 117543 Singapore, Singapore}
\affiliation{Department of Physics, National University of Singapore, Singapore 117542, Singapore}

\author{Christian Miniatura}
\affiliation{MajuLab, CNRS-UCA-SU-NUS-NTU International Joint Research Unit, Singapore}
\affiliation{Centre for Quantum Technologies, National University of Singapore, 117543 Singapore, Singapore}
\affiliation{Department of Physics, National University of Singapore, Singapore 117542, Singapore}
\affiliation{School of Physical and Mathematical Sciences, Nanyang Technological University, 637371 Singapore, Singapore}
\affiliation{Yale-NUS College, Singapore 138527, Singapore}
\affiliation{Universit\'{e} C\^{o}te d'Azur, CNRS, INPHYNI, Nice, France}

\date{\today}

\begin{abstract}
We demonstrate that a large class of one-dimensional quantum and classical exchange models can be described by the same type of graphs, namely Cayley graphs of the permutation group. Their well-studied spectral properties allow us to derive crucial information about those models of fundamental importance in both classical and quantum physics, and to completely characterize their algebraic structure. Notably, we prove that the spectral gap can be obtained in polynomial computational time, which has strong implications in the context of adiabatic quantum computing with quantum spin-chains. This quantity also characterizes the rate to stationarity of some important classical random processes such as interchange and exclusion processes. Reciprocally, we use results derived from the celebrated
Bethe ansatz to obtain original mathematical results about these  graphs in the unweighted case. We also discuss extensions of this unifying framework to other systems, such as asymmetric exclusion processes --- a paradigmatic model in non-equilibrium physics, or the more exotic non-Hermitian quantum systems.
\end{abstract}

\maketitle

Consider the following situations: (a) A card shuffling, where at each step, two randomly chosen adjacent cards of the deck are being switched; (b) a cold atom experiment involving strongly interacting ${}^{173}\mathrm{Yb}$ atoms confined in one dimension; (c) the quantum Heisenberg $XXX$ spin-chain; (d) the protein synthesis on RNA. What do these situations have in common? They can all be described by one-dimensional  exchange models, where the action of the Hamiltonian in the quantum case or of the transition matrix in the classical stochastic case is to exchange two adjacent elements \cite{Diaconis,Volosniev2014,Deuretzbacher2014,Heisenberg1928,MacDonald1969}. In fact, one of the purposes of this article is to show that they are all described by the {\em same} theoretical object, namely a graph associated with the permutation group. 

One-dimensional (1D) exchange models are ubiquitous in both quantum and classical physics, and their study has been associated with important theoretical breakthroughs. It was in particular to solve the homogeneous 1D Heisenberg spin-chain, a model of fundamental importance for the study of quantum magnetism \cite{Mattis1981}, that Bethe introduced his celebrated ansatz in 1931 \cite{Bethe1931}. His powerful insight, which attracted little attention at first, turned out to be one of the most fruitful theoretical achievements of the last century, as it has been extended and applied to a wide range of quantum \cite{Lieb1963,Yang1967,Sutherland1968,Andrei1983,Ambjorn2006} and classical \cite{Baxter1971,Batchelor1995,Golinelli2006} models that are said to be quantum integrable \cite{Sutherland2004}. However, extracting any physically relevant information from this method is still strenuous, and more importantly, the Bethe ansatz can no longer be applied if the system is inhomogeneous, as one could expect in a realistic experimental situation \cite{Korepin1993}. In this case, the study of such strongly correlated systems is greatly challenged by their computational complexity.

% In a recent work \cite{Decamp2020}, we have established that graph theory is a powerful framework in order to describe one peculiar type of exchange model, namely strongly repulsive mixtures of fermions confined in inhomogeneous continuous 1D potentials with open boundary conditions (OBC). This allowed us to completely characterize the algebraic structure of the system for any number of particles and any mixture, by obtaining the decomposition of the problem according to its exchange symmetries. Consequently, we derived explicit results about the spectrum of the system, that would have otherwise been impossible to compute in practice. 

% This unifying framework goes beyond the Bethe ansatz, since it allows us to describe inhomogeneous models.
% The results are similar to the ones obtained in Ref. \cite{Decamp2020}, with a notable exception in the ferromagnetic case, for which

In this article, we develop a universal graph-theoretical description in order to understand 1D exchange models containing inhomogeneities, with both periodic and open boundary conditions (respectively PBC and OBC). For its central importance, a peculiar focus is given to the inhomogeneous Heisenberg spin-chain and its generalization to fermionic spin-chains with spins greater than $1/2$ \cite{Sutherland1975}. This powerful framework allows us to derive crucial results about the system, which are true for any system size, number of spin-orientations, and spin configurations:
\begin{itemize}
\item We completely identify the algebraic structure of the system (i.e., its decomposition according to the irreducible representation of the permutation group). The following results are consequences of this fundamental decomposition.
\item We demonstrate a generalized Lieb-Mattis theorem \cite{LiebMattisPR} by identifying how the energy levels are ordered as a function of their symmetries in both the ferromagnetic and anti-ferromagnetic cases.
\item In the ferromagnetic case, we show how the energy gap between the ground-state and first-excited-state can be obtained in polynomial computational time. This huge computational advantage have important consequences in the field of adiabatic quantum computing, where the speed of the process is limited by the value of the energy gap -- which is a priori an exponentially-hard quantity to compute in the strongly correlated systems involved \cite{Albash2018}. 
\end{itemize}

The second important claim of this work is that the integrable case where the Bethe ansatz may be applied corresponds, within our graph-theoretical framework, to the {\em peculiar case} where the graphs are unweighted. As a consequence, we show that results derived from the Bethe ansatz may be applied in order to study the spectra of Cayley graphs of the permutation group with large number of elements. This approach should attract the attention of mathematicians. 

Furthermore, we discuss how our graph-theoretical framework and its generalizations may be applied to study a large class of quantum and classical 1D exchange models. These include the Fermi-Hubbard model and its non-Hermitian generalization for the quantum part, as well as 1D stochastic processes for the classical part, such as the interchange process and symmetric and asymmetric exclusion processes. The latter are paradigmatic in the context of non-equilibrium statistical physics \cite{Golinelli2006}. 

Some of the ground work for our results were first laid in Ref. \cite{Decamp2020} for a peculiar type of exchange model, namely strongly repulsive mixtures of fermions confined in inhomogeneous continuous 1D potentials with OBC. As further explained in the main text, the results we present here are significantly stronger and universal.

\section{Graph-theoretical description of inhomogeneous fermionic spin-chains} \label{secxxx}

\subsection{An exchange model}

\begin{figure}
	\includegraphics[width=1\linewidth]{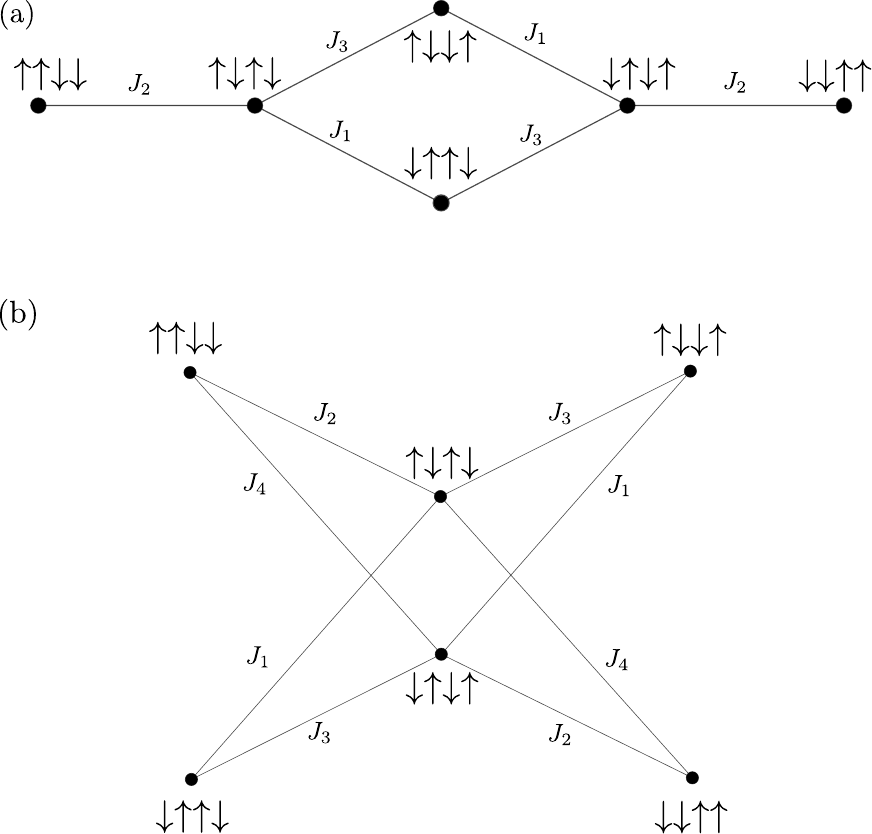}
	\caption{\label{ex4}Graphs describing a Heisenberg spin-chain $(2,2)$ of 2 spin-up and 2 spin-down. Panel (a): Open boundary conditions (OBC). Panel (b): Periodic boundary conditions (PBC). The graph with OBC can be obtained from the graph with PBC by removing the edges with weight $J_4$, corresponding to exchanges between distinguishable spins in positions $1$ and $4$. One can easily check that the Laplacian matrices of these graphs are equal to $\Delta^{(2,2)}$ (Eq.~\eqref{delta} of the main text).}
\end{figure}

We consider a system of $N$ spin-$S$, $S$ being a half-integer, on a 1D closed (PBC) or open (OBC) chain. We suppose that the populations in each spin component are fixed and given by a partition $\nu=(N_1,\ldots,N_{\kappa})$ of $N$, i.e., such that $N_1\ge\cdots\ge N_{\kappa}$ and $N_1+\cdots + N_{\kappa}=N$, with $\kappa\in\{2,\ldots,2S+1\}$. The Hamiltonian of the model is
\begin{equation}
\label{hamexch}
\mathcal{H}=\sum_{k=1}^{\tilde{N}}J_k\left(~\Id-P_{k,k+1}\right),
\end{equation}
where the local interaction constants $J_k$ verify either $J_k>0$ (respectively $J_k<0$) for all $k$ in the ferromagnetic (respectively antiferromagnetic) case, and operator $P_{k,k+1}$ exchanges the spin orientations at positions $k$ and $k+1$, e.g.,
\begin{equation}
P_{3,4}\ket{\uparrow\uparrow\downarrow\uparrow\downarrow\uparrow}=\ket{\uparrow\uparrow\uparrow\downarrow\downarrow\uparrow}.
\end{equation}
We have used the convention $\tilde{N}=N$ and $N+1=1$ for PBC and $\tilde{N}=N-1$ for OBC. Equivalently, we could have studied the following energy-shifted Hamiltonian:
\begin{equation}
\label{hamexchsut}
\tilde{\mathcal{H}}=-\sum_{k=1}^{\tilde{N}}J_kP_{k,k+1},
\end{equation}
but our graph-theoretical description will be more straightforward using Eq.~\eqref{hamexch}. A complete study of the Hamiltonian in Eq.~\eqref{hamexchsut} for the homogeneous case $J_1=\cdots=J_{\tilde{N}}=1$ using the Bethe ansatz is due to Sutherland \cite{Sutherland1975}. 

As first observed by Dirac \cite{Dirac1958}, in the case where $S=1/2$, $\mathcal{H}$ is equivalent to the Heisenberg spin-chain. Indeed, one can write $P_{k,k+1}$ as
\begin{equation}
\begin{split}
P_{k,k+1}&=\frac{\Id+\sigma_k^z\sigma_{k+1}^z}{2}+\sigma_k^+\sigma_{k+1}^-+\sigma_k^-\sigma_{k+1}^+\\
&=\frac{1}{2}(~\Id+\vec{\sigma}_k\cdot\vec{\sigma}_{k+1}),
\end{split}
\end{equation}
where we have used the standard notations for the Pauli matrices $\vec{\sigma}=(\sigma^x,\sigma^y,\sigma^z)$ and the raising and lowering operators $\sigma^{\pm}$ acting at positions $k$ and $k+1$. Therefore, the Heisenberg Hamiltonian of the $XXX$ spin-chain verifies
\begin{equation}
\label{hxxx}
\mathcal{H}_{XXX}=-\sum_{k=1}^{\tilde{N}}J_k~\vec{\sigma}_k\cdot\vec{\sigma}_{k+1}= 2\mathcal{H}-\sum_{k=1}^{\tilde{N}}J_k~\Id.
\end{equation}
In other words, the Heisenberg model is, up to an affine transformation, an exchange model. More generally, for spins larger than $1/2$, $P_{k,k+1}$ can always be written (up to an affine transformation) as a scalar product $\vec{\gamma}_k\cdot\vec{\gamma}_{k+1}$, where $\vec{\gamma}$ is the vector of all the generators $(\gamma^i)_{1\le i \le \kappa^2-1}$ of the $SU(\kappa)$ algebra \cite{Pfeifer2003}. In this sense, $\mathcal{H}$ can be seen as a $S>1/2$ generalization of the Heisenberg spin-chain for $S=1/2$.

For a given mixture $\nu=(N_1,\ldots,N_{\kappa})$, the number of distinguishable spin configurations along the chain is given by the following multinomial coefficient:
\begin{equation}
\label{dnu}
D_{\nu}=\frac{N!}{N_1!\cdots N_{\kappa}!}.
\end{equation}
\onecolumngrid

\begin{figure}
	\includegraphics[width=0.95\linewidth]{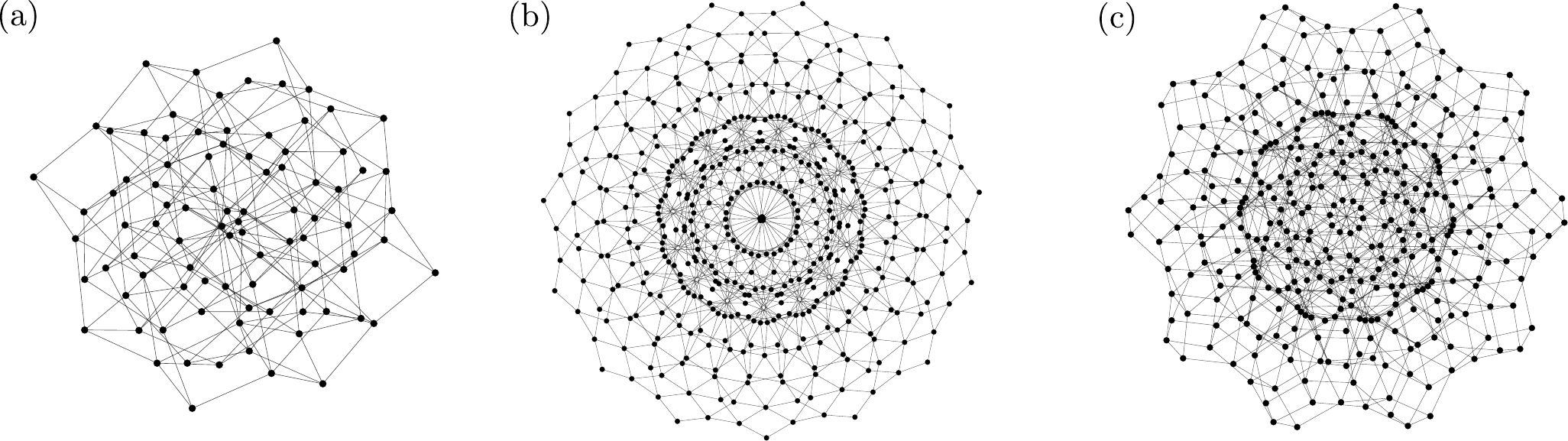}
	\caption{\label{exgr}Examples of graphs corresponding to different spin mixtures on closed chains (PBC). Panel (a): (2,2,2), panel (b): (12,3), panel (c): (7,2,1). These graphs display remarkable symmetry properties,  with a characteristic rose shape that is reminiscent of the boundary conditions.}
\end{figure}

\twocolumngrid 
Writing as $\ket{\chi}$ the quantum state associated with a given spin configuration $\chi$, each eigenvector $\ket{\psi}$ of $\mathcal{H}$ can be written $\ket{\psi}=\sum_{\chi}a_{\chi}\ket{\chi}$, with $\vec{a}=(a_{\chi})_{\chi}$ a vector of length $D_{\nu}$. Within this representation, $\mathcal{H}$ is a $D_{\nu}\times D_{\nu}$ real symmetric matrix $\Delta^{\nu}$ defined as
\begin{equation}
\label{delta}
\Delta^\nu_{\chi\chi'}=\left\{
\begin{array}{lll}
-J_k & \mbox{if } \ket{\chi} =P_{k,k+1} \ket{\chi'}\ne\ket{\chi'}\\ 
\sum_{k\in K_{\chi}}J_k& \mbox{if  } \ket{\chi}=\ket{\chi'}\\ 
0 & \mbox{ otherwise}
\end{array}
\right.,
\end{equation}
where $K_{\chi}=\{k~:~P_{k,k+1} \ket{\chi}\ne\ket{\chi}\}$.

As it is typically the case for strongly correlated systems, the dimension of the problem grows exponentially with the size $N$ of the system, and brute-force diagonalization becomes intractable even for moderate values of $N$. A comprehensive framework is thus necessary in order to overcome complexity and make general conclusions.

\subsection{Interpretation in terms of graph theory}

We suppose that the reader is familiar with basic notions of graph and group theories. Relevant definitions and properties as well as common references are included in the Appendix. 

We recall here the definitions of Cayley and Schreier graphs \cite{Brouwer2012}. Let $G$ be a finite group, and $S$ a generating subset of $G$  \cite{notegenerating} such that $s\in S$ if and only if $s^{-1}\in S$ ($S$ is said to be symmetric). The Cayley graph $X(G,S)$ associated with $G$ and $S$ is the graph whose vertices are indexed by all the elements of $G$, and such that there is an edge between $g$ and $g'$ if there is $s\in S$ such that $g'=sg$. The fact that $S$ is generating and symmetric ensures that $X(G,S)$ is connected and undirected (respectively). Additionally, if we consider a subgroup $H$ of $G$, the Schreier graph $X(H\subset G, S)$ is defined by indexing the vertices by all the cosets $gH$, and such that there is an edge between $gH$ and $g'H$ if there is $s\in S $ such that $g'H=sgH\ne gH$. The fact that $g'H\ne gH$ ensures that there are no self-loops. Note that if $H$ is the trivial subgroup, $X(H\subset G, S)\cong X(G,S)$. Besides, one could have easily defined the weighted versions of $X(G,S)$ and $X(H\subset G,S)$ by associating each $s\in S$ to a weight $w_s\in\mathbb{R}$.

Then, our graph-theory interpretation of Hamiltonian $\mathcal{H}$ and of its representation $\Delta^{\nu}$ is the following: $\Delta^{\nu}$ is the Laplacian matrix of the weighted Schreier graph $X(\mathfrak{S}_{\nu}\subset \mathfrak{S}_N,\tilde{S}_C)$, where:
			\begin{itemize}
\item $\mathfrak{S}_N$ is the permutation group of $\{1,\ldots,N\}$;
				\item $\mathfrak{S}_{\nu}=\mathfrak{S}_{N_1}^{\nu}\times\cdots\times\mathfrak{S}_{N_{\kappa}}^{\nu}$ is the Young subgroup associated with $\nu$, with $\mathfrak{S}_{N_i}^{\nu}$ the set of permutations $P\in\mathfrak{S}_N$ such that $P(i)=i$ if $i$ does not belong to $\{N_1+\cdots +N_{i-1}+1,\dots,N_1+\cdots + N_{i-1}+N_i\}$;
				\item $\tilde{S}_C\equiv\{(1,2),(2,3),\ldots,(\tilde{N},\tilde{N}+1)\}$ is the set of nearest-neighbour transpositions (Coxeter generators), where each $(k,k+1)\in \tilde{S}_C$ is associated with a weight $J_k$.				
			\end{itemize}
Note that the only difference between OBC and PBC is the presence of transposition $(N,1)$ in the PBC case. As a pedagogical example, the graphs associated with a $(2,2)$ mixture of two spin-up and two spin-down with both OBC and PBC are displayed in Fig.~\ref{ex4}. Other examples of graphs for more complex spin mixtures with PBC are given in Fig.~\ref{exgr}.

\subsection{Results}
\label{resg}

A first trivial observation is that, as it is well known for Laplacian matrices, the spectrum $\sigma\left(\Delta^{\nu}\right)$ of $\Delta^{\nu}$ is given by real positive numbers and contains $0$ with multiplicity $1$. Indeed, the multiplicity of $0$ in the spectrum of the Laplacian matrix is equal to the number of connected components of the graph, and here $X(\mathfrak{S}_{\nu}\subset \mathfrak{S}_N,\tilde{S}_C)$ is connected since $\tilde{S}_C$ generates $\mathfrak{S}_N$ (see Appendix \ref{apbasicgraph}). Therefore, the ground-state of $\mathcal{H}$ in the ferromagnetic case is non-degenerate with $0$ energy.

\paragraph{Algebraic structure.} 

One great advantage of noting that a graph is a Cayley or Schreier graph is that it allows one to access the algebraic structure of its spectrum in a natural way. Indeed, by associating a matrix to a graph that is itself associated with a group, we associate a matrix to a group: In other words, we are using the well-studied language of linear representations \cite{Fulton2004,Liebeck,Hamermesh_book}. In our case, for a given spin mixture $\nu$, we observe that each spin configuration is equivalent to a unique tabloid of shape $\nu$ \cite{Noteirrep}. The identification is done as follows: First, we associate each spin configuration $\ket{\chi}$ to a tabloid of shape $\nu$.  Then, we associate each row $i\in\{1,\ldots,\kappa\}$ to one spin orientation, and  we identify each index of row $i$ as the position of a spin of type $i$ along the chain. For instance, consider a mixture $(5,3)$ of $5$ spin-up and $3$ spin-down. Then, the following spin configuration and tabloid are equivalent:
\ytableausetup{tabloids,centertableaux}
\begin{equation}
\ket{\uparrow\downarrow\uparrow\uparrow\uparrow\downarrow\uparrow\downarrow}~~\Leftrightarrow~~\ytableaushort{13457,268}~.
\end{equation}
Therefore, $\Delta^{\nu}$ is written in the vector space whose basis is indexed by the tabloids, namely the permutation module $M^{\nu}$. More precisely, it can be written as:
\begin{equation}
\begin{split}
	\Delta^{\nu}&=\sum_{k=1}^{\tilde{N}}J_k\left(M^{\nu}(\mathrm{Id})-M^{\nu}((k,k+1))\right)\\
	&=M^{\nu}\left(\sum_{k=1}^{\tilde{N}} J_k\left(\mathrm{Id}-(k,k+1)\right)\right),
\end{split}
\end{equation}
where we have linearly extended $M^{\nu}$ to an algebra representation and $\mathrm{Id}$ and $(k,k+1)$ are respectively the identity and transposition in positions $k,k+1$ of $\mathfrak{S}_N$. Note that we just have re-written Eq.~\eqref{hamexch} in terms of representation theory. The interest of such a re-writing is that the decomposition of $M^{\nu}$ according to the irreducible representations (irreps) $S^{\mu}$ of $\mathfrak{S}_N$ (where the partition $\mu$ of $N$ labels uniquely the irrep) is well known and given by the so-called Young's rule \cite{James1984}:
\begin{equation}
\label{youngsrule}
M^\nu\cong\bigoplus_{\mu\trianglerighteq\nu}k_{\mu\nu}S^{\mu}\quad\quad (k_{\nu\nu}=1).
\end{equation}
Here, the multiplicities $k_{\mu\nu}$ are positive integers known as the Kostka numbers, and $\trianglerighteq$ is the dominance order over the set of partitions of $N$, which states that $\left[\mu_1,\ldots,\mu_r\right]\trianglerighteq\left[\nu_1,\ldots,\nu_r\right]$ if $\mu_1+\cdots+\mu_k\ge\nu_1+\cdots+\nu_k$ for all $k$. Thus, we obtain
\begin{equation}
\label{symdec}
\Delta^{\nu}\cong\bigoplus_{\mu\trianglerighteq\nu}k_{\mu\nu}S^{\mu}\left(\sum_{k=1}^{\tilde{N}} J_k\left(\mathrm{Id}-(k,k+1)\right)\right),
\end{equation}
with $k_{\nu\nu}=1$.
Eq.~\eqref{symdec} is a central result. Translated into words, it means that $\Delta^{\nu}$ is block-diagonal, and contains $k_{\mu\nu}$ identical blocks of size $\dim S^{\mu}$ corresponding to the irreps $S^{\mu}$ where $\mu\trianglerighteq\nu$. The fact that it does not contain irreps $S^{\mu'}$ such that $\nu\triangleright\mu'$ can be interpreted as a consequence of the Pauli principle. 

\paragraph{Symmetry ordering.}

Eq.~\eqref{symdec} implies that, given two spin mixtures $\nu$ and $\nu'$,
\begin{equation}
\label{specinc}
	\nu\trianglerighteq\nu'\quad\Rightarrow\quad\sigma\left(\Delta^{\nu}\right)\subseteq\sigma\left(\Delta^{\nu'}\right).
\end{equation}
In particular, the spectral radii (the largest absolute value of the eigenvalues) verify $\rho\left(\Delta^{\nu}\right)\le\rho\left(\Delta^{\nu'}\right)$. Importantly, one can then show the important fact that it is always possible to construct an eigenvector $\vec{a}_{\nu}$ with eigenvalue $\rho\left(\Delta^{\nu}\right)$ that belongs to the symmetry class $[\nu]$ (in other words, which is ``as anti-symmetric as possible''). The proof of this statement is exactly the same as the one we described in Ref. \cite{Decamp2020}: First we note that the graph $X(\mathfrak{S}_{\nu}\subset \mathfrak{S}_N,\tilde{S}_C)$ is bipartite and can be separated into even and odd spin configurations $\ket{\chi}$, then we use this fact to construct a vector that belongs to the symmetry class $[\nu]$ and show that it belongs to the eigenspace with eigenvalue $\rho\left(\Delta^{\nu}\right)$. Note that the fact that $k_{\nu\nu}=1$ in Eq.~\eqref{symdec} implies that $\vec{a}_{\nu}$ is in fact unique for generic choices of weights $(J_k)_k$ (see also \cite{Poignard2018}). In the context of Ref. \cite{Decamp2020}, the fact that the spectrum of $\Delta^{\nu}$ has a negative contribution to the energy implies that $\rho\left(\Delta^{\nu}\right)$ corresponds to the ground-state of a given mixture $\nu$. Therefore, the fact that this ground-state belongs to the symmetry class $[\nu]$ has been interpreted as a generalization of the Lieb-Mattis theorem \cite{LiebMattisPR}. Here, the ground-state energy is $-\rho\left(\Delta^{\nu}\right)$ for the antiferromagnetic case $J_k<0$, and we can reach a conclusion similar to Ref.~\cite{Decamp2020}. Nevertheless, the situation is different for the ferromagnetic case $J_k>0$, where $\rho\left(\Delta^{\nu}\right)$ corresponds to the maximal energy of the mixture $\nu$, which is of course less physically interesting. It is however an important lemma for what follows. Indeed, Eq.~\eqref{specinc} also shows that all the spectra are included in the spectrum of $\Delta^{(1,\ldots,1)}\equiv \Delta$ corresponding to the case where $N=\kappa$. In this case, $X(\mathfrak{S}_{(1,\ldots,1)}\subset \mathfrak{S}_N,\tilde{S}_C)$ is isomorphic to the Cayley graph $X(\mathfrak{S}_N,\tilde{S}_C)$. It is a bipartite graph, since it can be split between even and odd permutations. It is also $d-$regular, with
\begin{equation}
\label{degree}
	d=\sum_{k=1}^{\tilde{N}}J_k.
\end{equation}
Therefore, one has
\begin{equation}
\label{symspec}
	\lambda\in\sigma\left(\Delta\right)\quad\Leftrightarrow\quad 2d-\lambda\in\sigma\left(\Delta\right).
\end{equation}
Indeed, if $\vec{a}_{\lambda}$ is an eigenvector of $\Delta$ with eigenvalue $\lambda$, it is easy to see that the vector $\tilde{a}_{\lambda}$ that coincides with $\vec{a}_{\lambda}$ for vertices corresponding to even permutations and is equal to $-\vec{a}_{\lambda}$ for odd vertices is an eigenvector of $\Delta$ with eigenvalue $2d-\lambda$. Moreover, if $\vec{a}_{\lambda}$ belongs to the symmetry class $[\mu]$, we observe that, by construction, $\tilde{a}_{\lambda}$ belongs to the conjugate symmetry class $[{}^t\mu]$. For instance, $0\in\Delta$ belongs to the trivial representation $[N]$ (totally symmetric) and $2d=\rho(\Delta)$ belongs to the sign representation $[1,\ldots,1]$ (totally anti-symmetric). More generally, if we denote by $E([\mu])$ the lowest eigenvalue of $\Delta$ that belongs to the symmetry class $[\mu]$, we deduce from the previous discussion that, with both OBC and PBC, the following generalized Lieb-Mattis theorem holds:
\begin{equation}
\mu\trianglerighteq\mu'\quad\Rightarrow\quad \left\{
\begin{array}{ll}
E([\mu])\le E([\mu']) & \mbox{if } J_k>0 \\ 
E([\mu])\ge E([\mu']) & \mbox{if } J_k<0
\end{array}
\right..
\end{equation}
Thus, we observe opposite behavior in the ferromagnetic and antiferromagnetic phases, where the ground-state tends to be more symmetric in the first case and more anti-symmetric in the second case.

\paragraph{Energy gap.}

\begin{figure}
	\includegraphics[width=0.7\linewidth]{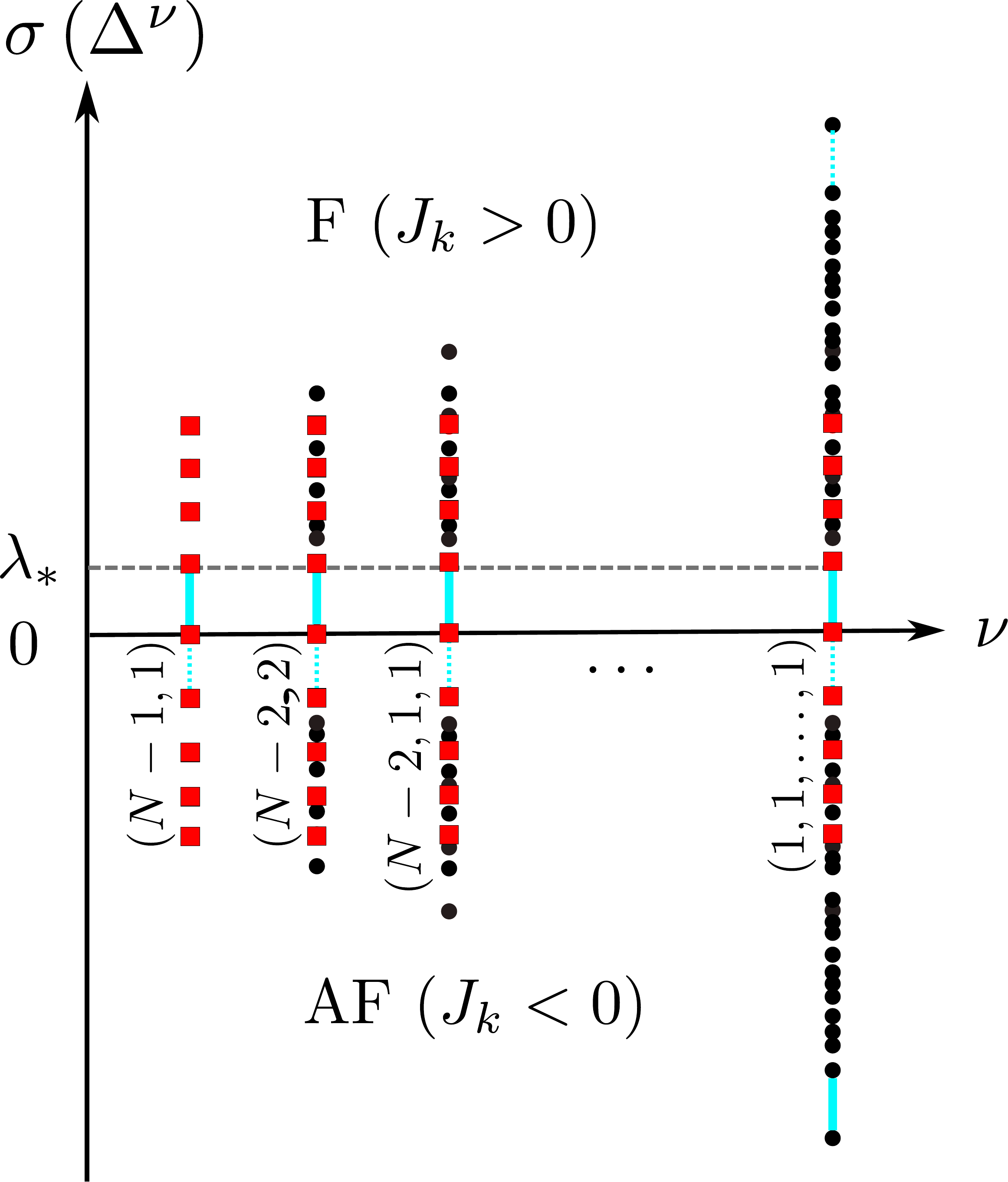}
	\caption{\label{figgap}Spectra $\sigma\left(\Delta^{\nu}\right)$ of the Laplacian matrices $\Delta^{\nu}$ for different spin mixtures $\nu$, where $\nu$ is a partition of $N$, in both the ferromagnetic ($F$, $J_k>0$) and antiferromagnetic ($AF$, $J_k<0$) cases. In the $F$ phase, the energy gap $\lambda_*$ of the simplest non-trivial case $(N-1,1)$ is the same as the energy gap of {\em any} mixture $\nu$, which gives a huge computational advantage. This is no longer true in the $AF$ phase, where the spectra are opposite as compared to the $F$ phase, so that $\lambda_*$ is associated with the energy differences between the most and second-most excited states. The only exception is in the case where $\nu=(1,1,\ldots,1)$ (i.e., $N=\kappa$), due to the symmetry of the spectrum in this case (Eq.~\eqref{symspec} of the main text). $\lambda_*$ is represented by a solid (respectively dashed) light blue line when associated with the energy gap (respectively the gap between the second-most and most excited states). The spectrum of $\Delta^{(N-1,1)}$, which is included in $\sigma\left(\Delta^{\nu}\right)$ for all mixture $\nu$ due to Eq.~\eqref{specinc}, is represented by red squares, and other eigenvalues by black dots.}
\end{figure}

Let us discuss the ferromagnetic case, as the antiferromagnetic is similar to what we described in Ref. \cite{Decamp2020}. If we denote the spectral gap of $\Delta$ by $\lambda_*$, which in this case is the smallest non-zero eigenvalue, the fact that $\dim S^{[N]}=1$, together with the fact that $0$ has multiplicity of $1$ and that $[N-1,1]\trianglerighteq\mu$ for any partition $\mu\ne [N]$ implies:
\begin{equation}
\label{gap}
\lambda_*=E([N-1,1]).
\end{equation}
Besides, it is easy to deduce from Eq.~\eqref{symdec} that $\lambda_*$ is in fact the spectral gap of $\Delta^{\nu}$ for any $\nu\ne (N)$. In other words, for any spin mixture, the energy gap is the same as in the polaron case with $N-1$ spin-up and 1 spin-down. It is therefore sufficient to compute the lowest non-zero eigenvalue of the following $N\times N$ matrix:
\begin{equation}
\label{matpath}
\left(\begin{smallmatrix}
J_1 + J_N & -J_1  &  & & -J_N \\
-J_1& J_1+J_2 & -J_2  & & \\
& \ddots & \ddots & \ddots & \\
& & -J_{N-2} & J_{N-1}+J_{N-2} & -J_{N-1}\\
-J_N & &  & -J_{N-1} & J_{N-1}+J_N
\end{smallmatrix}\right),
\end{equation}
where $J_N=0$ in the OBC case. Depending on the boundary conditions, Eq.~\eqref{matpath} is either the Laplacian matrix $\Delta^{(N-1,1)}$ of a cycle graph $C_N$ (PBC, here e.g., for $N=9$):
\vspace{0.2cm}
\begin{center}
	\includegraphics[width=0.5\linewidth]{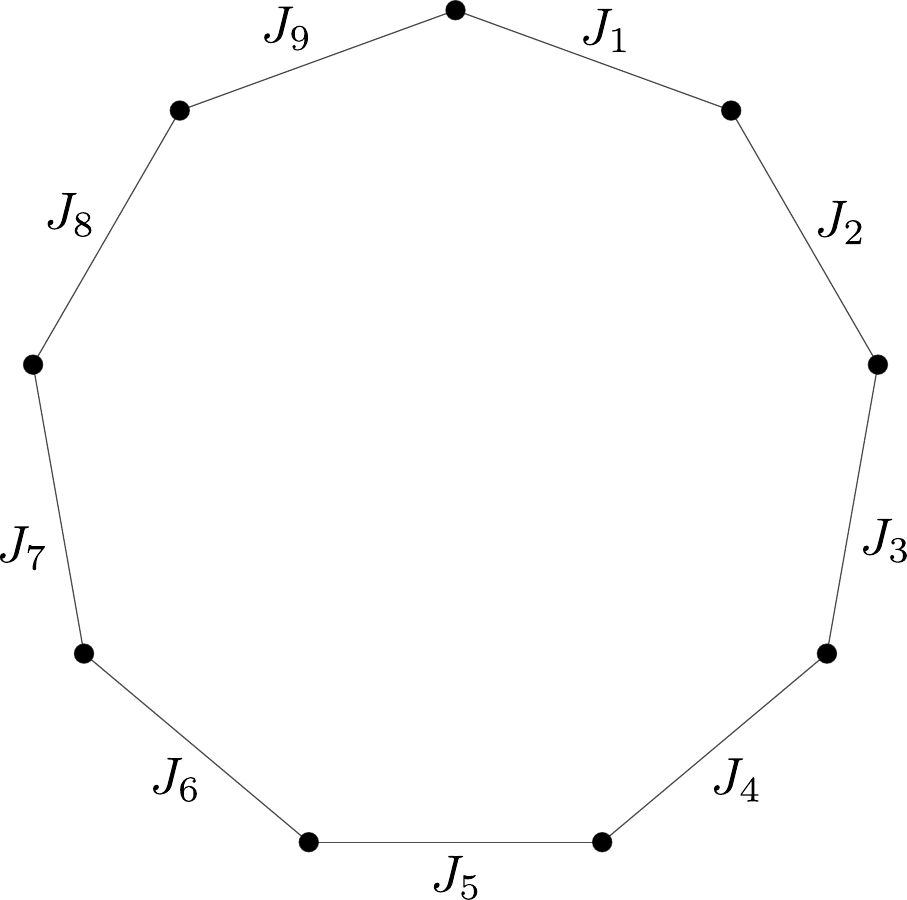}
\end{center}
\vspace{0.2cm}
or a path graph $P_N$ (OBC, here e.g., for $N=9$):
\vspace{0.2cm}
\begin{center}
	\includegraphics[width=0.98\linewidth]{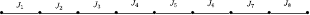}
\end{center}
\vspace{0.2cm}
This considerably reduces the complexity of the problem (typically from $N!$ to $N$). Note that Bacher has given an alternative proof of Eq.~\eqref{gap} for the unweighted Cayley graph $X(\mathfrak{S}_N,S_C)$ with $S_C=\{(1,2),(2,3),\ldots,(N-1,N)\}$ in a purely mathematical context \cite{Bacher1994}. Here, our result relies only on symmetry arguments. We also remark that in Ref. \cite{Decamp2020}, the fact that the Laplacian spectrum has a negative contribution to the energy implies that one can only use Eq.~\eqref{gap} in the case $N=\kappa$ where Eq.~\eqref{symspec} is valid. Here, in the ferromagnetic case, the result is much stronger, since one can use this method for any spin mixture $\nu$. This result is illustrated in Fig.~\ref{figgap}.

Thus, the application of these results to adiabatic quantum computing, where the speed of the process depends crucially on the value of the energy gap, are extremely promising \cite{Albash2018}. For instance, with 100 qubits with 50 spin-up and 50 spin-down that are subjected to $\mathcal{H}$, one only has to diagonalize the  $\Delta^{(99,1)}$ matrix of size 100$\times$100 in order to get the spectral gap, instead of the whole $\Delta^{(50,50)}$ matrix of size $D_{(50,50)}\times D_{(50,50)}\approx 10^{29}\times 10^{29}$. A subsequent question for future work is to characterize the class of problems that can be studied by an adiabatic tuning of the exchange coefficients $J_k$ in Hamiltonians of the form $\mathcal{H}$ (Eq.~\eqref{hamexch}).

\section{Homogeneous case. Connection to the Bethe ansatz} \label{secba}

\subsection{Explicit eigenvalues}

We now consider the homogeneous case where $J_1=\cdots=J_N=1$. In this situation the spectrum of the path graph $P_N$ and cycle graph $C_N$ are well-known and are respectively given by \cite{Anderson1985}:
\begin{equation}
\label{specpath}
\sigma\left(P_N\right)=\left\{2-2\cos\left(\frac{j\pi}{N}\right),~ 0\le j \le N-1\right\},
\end{equation}
and
\begin{equation}
\label{speccycle}
	\sigma\left(C_N\right)=\left\{2-2\cos\left(\frac{2j\pi}{N}\right),~ 0\le j \le N-1\right\}.
\end{equation}
In particular, Eq.~\eqref{gap} implies that, for any mixture $\nu$, we get an explicit expression for the energy gap:
\begin{equation}
\label{gaphom}
\lambda_*=2-2\cos\left(\frac{b\pi}{N}\right),
\end{equation}	
with $b=1$ in the OBC case and $b=2$ in the PBC case.		  
Eq.~\eqref{gaphom} has been checked for several examples, such as the ones provided in Fig. \ref{exgr}. We observe that the energy gap vanishes as $(b \pi /N)^2$ in the thermodynamic limit $N\to \infty$. In terms of the aforementioned applications to adiabatic quantum computing, this means that the coupling constant $J_k$ should be strong (typically $J_k=O\left(N^2\right)$) in order to have a substantial energy gap.

\subsection{Bethe ansatz in the PBC case}

The homogeneous version of $\mathcal{H}$ (Eq.~\eqref{hamexch}) in the PBC case has been intensively studied. In fact, it is for the $S=1/2$ version of this model (cf Eq.~\eqref{hxxx}) that Bethe introduced in 1931 his celebrated ansatz, an educated guess about the form of the eigenvectors and eigenvalues of $\Delta^{(N_1,N_2)}$ \cite{Bethe1931}. His solution has been later generalized to the $S>1/2$ model by Sutherland \cite{Sutherland1975}, and to many other classes models, ranging from bosons and fermions in the one-dimensional continuum \cite{Lieb1963,Yang1967,Sutherland1968} to the $XYZ$ spin-chain and two-dimensional ice-type models \cite{Baxter1971} --- the so-called quantum integrable systems \cite{Sutherland2004}. This suggests to us that the graphs $X(\mathfrak{S}_{\nu}\subset\mathfrak{S}_N, \tilde{S}_C)$ we introduced and their weighted versions, which correspond to models that are no longer integrable, have an important role in a large class of strongly correlated systems. We mention some of them in the next Section.

Although it is not the purpose here to provide the details of this well-established theory, we find it useful to briefly recall Bethe's original solution to the $S=1/2$ XXX spin-chain, and its generalization to higher spins by Sutherland. This allows us to obtain results on the spectra of the unweighted graphs $X(\mathfrak{S}_{\nu}\subset\mathfrak{S}_N, \tilde{S}_C)$ that are, up to our knowledge, not familiar to mathematicians.

In the $S=1/2$ case where $\nu=(N_1,N_2)\equiv (N-M,M)$, any spin configuration $\ket{\chi}$ is characterized by the positions of the $M$ down spins, and one can write:
\begin{equation}
	\ket{\psi} = \sum_{1\le x_1 <\cdots < x_M\le N}a(x_1,\ldots,x_M)\ket{x_1,\ldots,x_M},
\end{equation}
where $x_i$ is the position of the $i^{\mathrm{th}}$ spin-down. If $M=1$, the Schr\"{o}dinger equation is easily solved, yielding the spectrum given in Eq.~\eqref{speccycle} with eigenvectors of the (Bloch) form $a(x_1)=A e^{ik_1x_1}$, where $A\in \mathbb{R}$ and $k_1=2j\pi/N$ ($0\le j \le N-1$). Then, Bethe's intuition was that in the general case $M\ge 1$, the $a(x_1,\ldots,x_M)$ coefficients could always be written as a finite sum of exponentials that is now widely known as the Bethe ansatz:
\begin{equation}
\label{betheansatz}
	a(x_1,\ldots,x_M)=\sum_{P\in \mathfrak S_M}A_P ~e^{i(k_{P1}x_1+\cdots+k_{PM}x_M)}.
\end{equation}
Then, one can show that if $a(x_1,\ldots,x_M)$ has this form and if the Schr\"{o}dinger equation is verified, one can relate $A_P$ for $P\in\mathfrak{S}_M$ with $A_Q$ for $Q=(P1,\ldots,P(j+1),Pj,\ldots,PM)=(j,j+1)\circ P$ by a simple phase factor:
\begin{equation}
\label{phasefactor}
	A_P=-e^{-i\theta(k_{Pj},k_{P(j+1)})}A_Q, 
\end{equation}
where 
\begin{equation}
	\theta(k_j,k_l)=2\arctan \frac{1}{2}\left(\cot \frac{k_j}{2}-\cot \frac{k_l}{2}\right).
\end{equation}
In order to determine the $M$ momentum-like quantities $k_1,\ldots,k_M$, the next step is to apply PBC by writing $a(x_1,\ldots,x_M)=a(x_2,\ldots,x_M,x_1+N)$, which yields
\begin{equation}
	e^{ik_jN}=(-1)^Me^{-i\sum_{l=1}^M\theta(k_j,k_l)} \quad\quad(1\le j \le M),
\end{equation}
or equivalently
\begin{equation}
\label{baeq}
	\left(\frac{\Lambda_j+i/2}{\Lambda_j-i/2}\right)^N=\prod_{l=1,l\neq j}^{M}\frac{\Lambda_j-\Lambda_l+i}{\Lambda_j-\Lambda_l-i}\quad\quad(1\le j \le M),
\end{equation}
where $\Lambda_j=\frac{1}{2}\cot \frac{k_j}{2}$.
The $M$ coupled non-linear equations of the form of Eq.~\eqref{baeq} are the celebrated Bethe ansatz equations. Then, Bethe showed that their solutions (with unequal $\Lambda_j$'s) yield indeed \textit{all} the eigenvectors of $\Delta^{(N-M,M)}$, with eigenvalues given by
\begin{equation}
E=\sum_{j=1}^M\frac{1}{\Lambda_j^2+1/4}.	
\end{equation}
Note that PBC are crucial in order to have the correct number of coupled equations.

Interestingly, the generalization to the $S>1/2$ case by Sutherland was performed 44 years after Bethe's original solution. The idea is to apply Bethe's ansatz successively $\kappa-1$ times: For a mixture $\nu=(N_1,N_2\ldots,N_{\kappa})$, we first separate the $N_1$ spins of type 1 with the $M_1=N_2+\cdots+N_{\kappa}=N-N_1$ others, and treat the spins of type 1 as the spin-up and the $M_1$ others as the spin-down of the previous example. Then, we separate the $M_1$ remaining spins between the $N_2$ spins of type 2 and $M_2=N_3+\cdots+N_{\kappa}=M_1-N_2$ others, that we treat respectively as the spin-up and spin-down of the $S=1/2$ case, and repeat the procedure. This technique is known as the nested Bethe ansatz. We will not write the resulting Bethe ansatz equations here, which are more involved but whose forms are very similar to the ones described above for $S=1/2$ (cf \cite{Sutherland1975} for more details).

Thus, the Bethe ansatz provides a method for diagonalizing the Laplacian matrix $\Delta^{\nu}$ of the unweighted Schreier graph $X(\mathfrak{S}_{\nu}\subset\mathfrak{S}_N, \tilde{S}_C)$. However, it should be noted that the non-linear coupled equations of the form Eq.~\eqref{baeq} are hard to solve in practice for even moderate values of $N$. However, in the thermodynamic limit where $N,N_1,\ldots,N_\kappa~\to\infty$ while keeping $N_j/N$ constant, one can write the logarithmic version of the Bethe ansatz equations as coupled integral (Fredholm) equations, allowing to derive some exact analytical results \cite{Sutherland2004}. For instance, translating a result from Sutherland for the ground-state energy per particle in the balanced case $\nu=(N/\kappa,\ldots,N/\kappa)$ in our graph-theoretical language, we get that the spectral radius of $\Delta^{(N/\kappa,\ldots,N/\kappa)}$ is given, in the large $N$ limit, by:
			\begin{equation}
			\label{digamma}
			\rho\left(\Delta^{(N/\kappa,\ldots,N/\kappa)}\right)\simeq -\frac{2N}{\kappa}\left(\Psi\left(\frac{1}{\kappa}\right)+\gamma\right),
			\end{equation}			  
where $\Psi$ is Euler's digamma function and $\gamma=\Psi(1)\simeq 0.577215$ is Euler's constant. In the large $\kappa$ limit, one also has
			\begin{equation}
			\rho\left(\Delta^{(N/\kappa,\ldots,N/\kappa)}\right)\simeq 2N\sum_{k=2}^{\infty}\frac{(-1)^k\zeta(k)}{\kappa^k},
			\end{equation}			  
with $\zeta(k)=\sum_{n\ge 1} 1/n^k$ the Riemann zeta function. We have checked Eq.~\eqref{digamma} by explicitly diagonalizing $\Delta^{(N/\kappa,\ldots,N/\kappa)}$ for different values of $N$ and $\kappa$. We found that this equation is already valid with less than a $1\%$ relative error for $N=12$.

To conclude this discussion on Bethe ansatz, we see that analytical results obtained from the theory of quantum integrable systems can be used to study the Laplacian spectra of Schreier graphs of the permutation group $\mathfrak{S}_N$ for large values of $N$. The approach here may potentially interest mathematicians working in the fields of asymptotic, combinatorial and geometric group theories.

\section{Other quantum and classical exchange models} \label{secom}

\subsection{Other models described by $X(\mathfrak{S}_{\nu}\subset\mathfrak{S}_N, \tilde{S}_C)$}

\begin{figure}
	\includegraphics[width=0.95\linewidth]{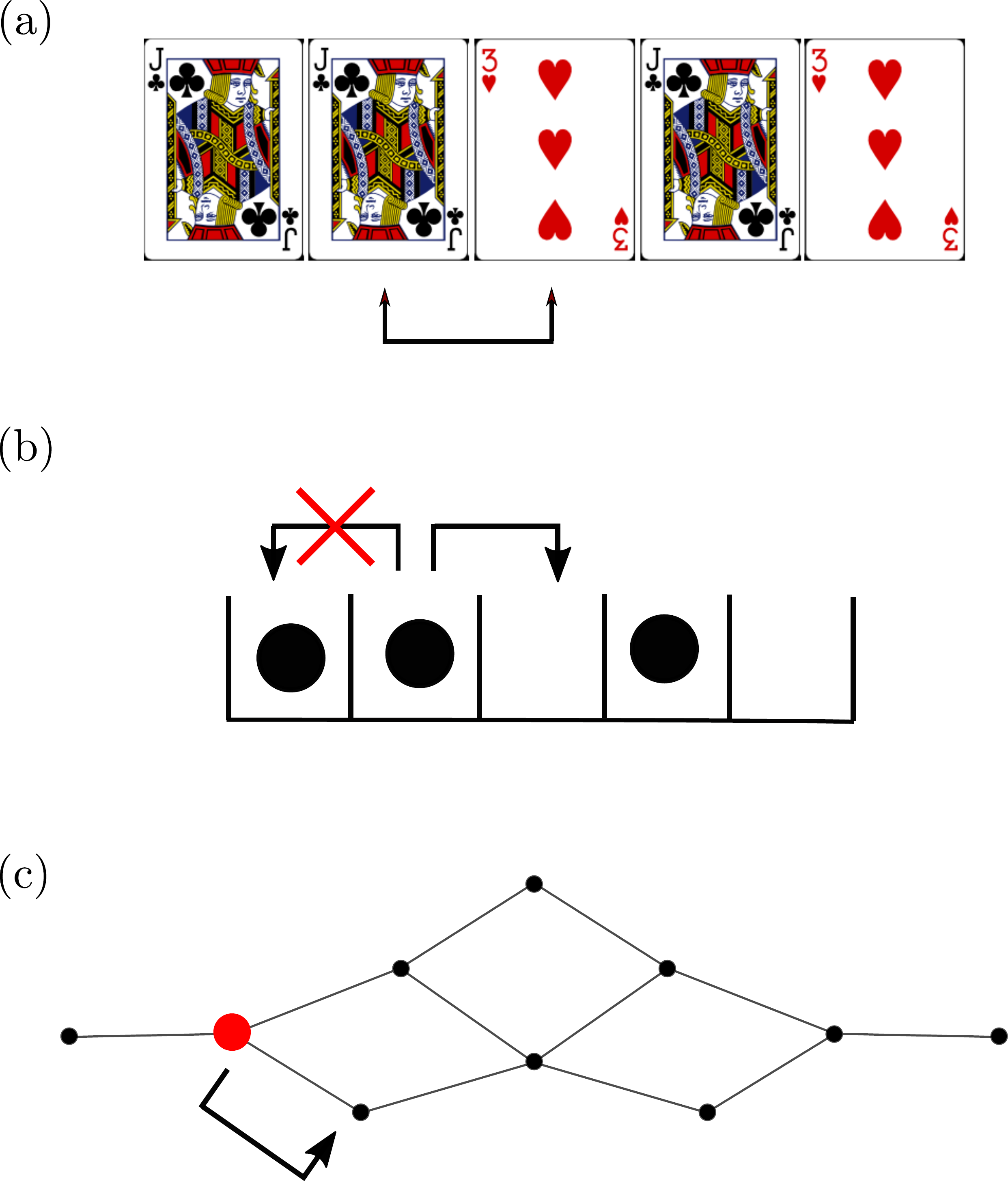}
	\caption{\label{deck} Three equivalent random processes. Panel (a): An interchange process with 3 black cards and 2 red cards. Panel (b): A symmetric exclusion process with 3 particles on 5 sites. Panel (c): A random walk on $X(\mathfrak{S}_{(3,2)}\subset\mathfrak{S}_5, \tilde{S}_C)$.}
\end{figure}

\paragraph{Quantum models.} So far we have focused on the Heisenberg model and its higher spin generalization. In Ref. \cite{Decamp2020}, we have shown that strongly repulsive $SU(\kappa)$ mixtures of ultracold fermions confined in 1D continuous potentials can also be efficiently described using the $X(\mathfrak{S}_{\nu}\subset\mathfrak{S}_N, \tilde{S}_C)$ graphs. This is due to a mapping between this model and a discrete spin-chain (the OBC version of $\mathcal{H}$, Eq.~\eqref{hamexch}), as initially observed in Refs. \cite{Volosniev2014,Deuretzbacher2014}. Subsequently, other mappings allow one to extend our graph description to other models.

A first example is the Hubbard chain, which can be mapped to the Heisenberg model using the well-known Jordan-Wigner transformation \cite{Jordan1928}. More precisely, let us consider a system consisting of $M$ spinless fermions on an $N-$site chain that is described by the following Hamiltonian:
\begin{equation}
\label{hubbard}
\begin{split}
\mathcal{H}_{\mathrm{F}}=-t&\sum_{k=1}^{\tilde{N}}\left[c_{k+1}^\dagger c_k+\mathrm{h.c.}\right]\\&+V\sum_{k=1}^{\tilde{N}}\left(c_{k+1}^\dagger c_{k+1}-\frac{1}{2}\right)\left(c_{k}^\dagger c_{k}-\frac{1}{2}\right),
\end{split}
\end{equation}
where $c_k^\dagger/c_k$ are the fermionic creation/annihilation at site $k$. Then, $\mathcal{H}_{\mathrm{F}}$ is equivalent to the homogeneous Heisenberg spin-chain $\mathcal{H}_{XXX}$ (Eq.~\eqref{hxxx}) with $J_1=\cdots=J_k=2t=V$. Therefore, it can be described by $X(\mathfrak{S}_{(N-M,M)}\subset\mathfrak{S}_N,\tilde{S}_C)$, and results derived in Section \ref{secxxx} can be applied here.

\paragraph{Classical models: A deck of cards.}Perhaps more surprisingly, this framework can also be applied to classical physics, in the context of random walks. Indeed, the Laplacian of a graph is related to the transition matrix of a random walk on this graph \cite{Lovász93randomwalks}. Here, a random walk on $X(\mathfrak{S}_{\nu}\subset\mathfrak{S}_N, \tilde{S}_C)$ can be interpreted as a so-called interchange process \cite{Diaconis}. For example, consider a deck of $N$ cards of types $\nu=(N_1,\ldots,N_{\kappa})$ ($N_i$ cards of type $i$), placed on an open or closed chain. The number of card configurations $\chi$ is $D_{\nu}$, as defined in Eq.~\eqref{dnu}. Then, at each step, a random pair of adjacent cards are selected and exchanged with a probability $J_k\in[0,1]$ such that $\sum_{k=1}^{\tilde{N}}J_k=1$. In the uniform case, one has $J_k=1/\tilde{N}$ for each $k$. The transition matrix $T^{\nu}$ of this Markov process, which specifies the probability of going from a configuration $\chi$ to a configuration $\chi'$, is given by
\begin{equation}
\label{tm}
T^{\nu}_{\chi\chi'}=  \left\{
      \begin{aligned}
        0\quad\quad\quad&\text{ if }\chi\ne(k,k+1)\chi'\\
        J_k\quad\quad~~&\text{ if }\chi=(k,k+1)\chi'\text{ and }\chi\ne\chi'\\
        1-d(\chi)&\text{ if }\chi=\chi'\\
      \end{aligned}
    \right.,
\end{equation}
where $d(\chi)$ is the degree of the vertex corresponding to $\chi$ in $X(\mathfrak{S}_{\nu}\subset\mathfrak{S}_N, \tilde{S}_C)$. Then, it is clear that $T^{\nu}$ is related to the Laplacian matrix $\Delta^{\nu}$ of $X(\mathfrak{S}_{\nu}\subset\mathfrak{S}_N, \tilde{S}_C)$ by
\begin{equation}
\label{tmd}
T^{\nu}=~\Id-\Delta_{\nu}.
\end{equation}
In particular, the eigenvalues $1=\beta_0>\beta_1\ge\cdots\ge\beta_{N-1}>-1$ of $T^{\nu}$ are trivially related to the spectrum of $\Delta^{\nu}$. As it is well-known in probability theory, the spectrum of a transition matrix of a Markov process is related to its rate to stationarity, or in simpler words in our case, to the speed at which a deck of cards can be considered as fully randomized by our interchange shuffle. Thus, the distance $
\|(T^{\nu})^m-\mu\|$ to the stationary measure $\mu$ after $m$ steps (see e.g., Refs. \cite{Diaconis,Diaconis1993,Ng1996} for standard definitions of the norm $\|\cdot\|$ and the stationary measure $\mu$) is related to $\beta_*=\max\left(\beta_1,|\beta_{N-1}|\right)$  by \cite{Diaconis1993}
\begin{equation}
\label{ratetos}
\|(T^{\nu})^m-\mu\|\le \frac{\sqrt{\tilde{N}}}{2}\beta_*^m.
\end{equation}
In Section \ref{secxxx}, we have shown that the spectral gap of $\Delta^{\nu}$ in the ferromagnetic case $J_k>0$ is equal to the spectral gap $\lambda_*$ of $\Delta{(N-1,1)}$ for every mixture $\nu$. Therefore, Eqs.~\eqref{tmd} and \eqref{ratetos} yields $\beta_1=1-\lambda_*$. In particular in the uniform case $J_1=\cdots=J_{\tilde{N}}=1/\tilde{N}$, using Eq.~\eqref{gaphom}, we have found
\begin{equation}
\label{beta1}
\beta_1=1-\frac{2}{\tilde{N}}+\frac{2}{\tilde{N}}\cos\left(\frac{b\pi}{N}\right).
\end{equation}
Moreover, one has $|\beta_{N-1}|=\rho(\Delta^{\nu})-1$, with $\rho(\Delta^{\nu})$ the spectral radius of $\Delta^{\nu}$, for which there is no analytical formula in the general case. For a uniform process on a closed chain, in the balanced case $N_1=\cdots=N_{\kappa}=N/\kappa$ and for sufficiently large values of $N$, one may use the Bethe ansatz result of Eq.~\eqref{digamma} and get
\begin{equation}
\label{betan}
|\beta_{N-1}|\simeq -\frac{2}{\kappa}\left(\Psi\left(\frac{1}{\kappa}\right)+\gamma\right)-1.
\end{equation}
Besides, one may also use one of the well-known bounds on the Laplacian spectral radii of graphs (see e.g., Eq.~\ref{biregmax}). More generally, all the results derived in the previous sections, such as Eqs.~\eqref{symdec} and \eqref{specinc}, or the generalized Lieb-Mattis theorem, can also be translated in this context. Thus, we see that our framework can also efficiently be used in order to analyse interchange processes.

It is important to note that in the case where the interchange process consists of $\kappa=2$ types of cards, it is equivalent to the so-called symmetric exclusion process, where $M$ particles are on a chain of $N$ sites, and at each step a particle is randomly selected and jumps either to the right or to the left if the corresponding site is empty \cite{notejw}. In this case, the link to the uniform Heisenberg model and the Bethe ansatz, which is clear with our graph description, have already been exploited by mathematicians in order to study some properties of these stochastic processes (see e.g., Ref. \cite{Ng1996}). The equivalence between interchange and exclusion processes as well as our graph description of these models are summarized in Fig \ref{deck}.

\subsection{Generalizations}

\begin{figure}
	\includegraphics[width=0.95\linewidth]{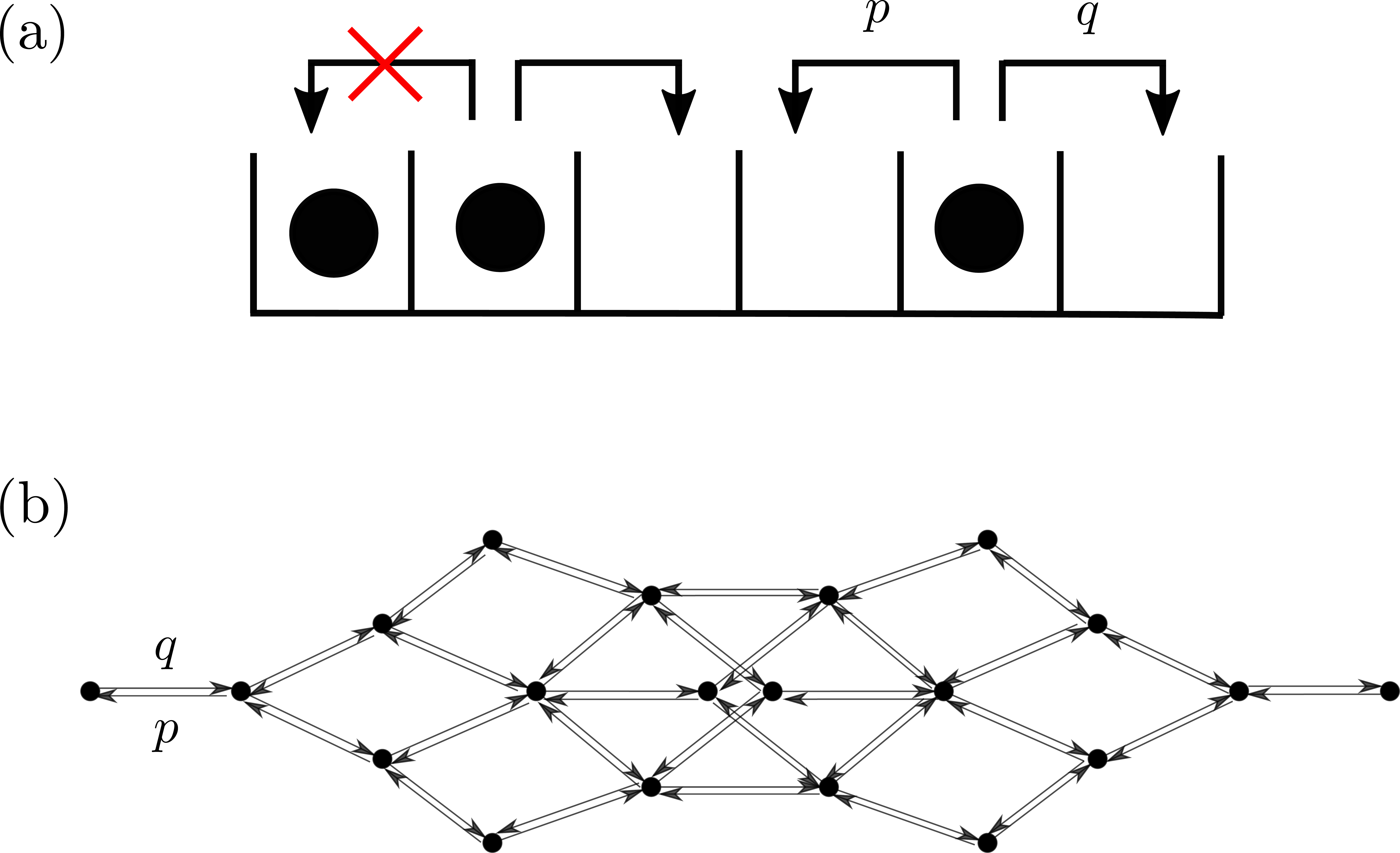}
	\caption{\label{asym} Panel (a): Illustration of an asymmetric exclusion process, with probability $p=1-q\in [0,1]$ and $p\ne q$. Alternatively, illustration of a non-Hermitian Hubbard model with asymmetric hopping. Panel (b): Directed graph $X_{\rightleftarrows}(\mathfrak{S}_{(3,3)}\subset\mathfrak{S}_6, \tilde{S}_C)$ associated with panel (a). The edges corresponding to a hopping to the left/right have a weight $p$/$q$ (respectively).}
\end{figure}

Finally, with some {\em ad hoc} modifications, it is possible to extend our graph-theoretical description to other models. We now proceed to briefly describe two examples of such models.

The first example are the so-called asymmetric exclusion processes. As its name suggests, it is the same as symmetric exclusion processes, except that the probability $p_k$ of a particle on site $k$ to jump on an empty site on the left  is different from the probability $q_k$ of jumping to the right  [Fig.~\ref{asym}(a)]. Such models are paradigmatic in the context of non-equilibrium statistical physics, and describe a wide range of phenomena, ranging from the protein synthesis on RNA \cite{MacDonald1969}, hopping conductivity in solid electrolytes \cite{Richards1977} and surface growth processes \cite{Krug1997}  to traffic flows \cite{Wolf1998} and molecular rotors \cite{Klumpp2003}. Similarly to the symmetric exclusion processes, these models are soluble with the Bethe ansatz in the uniform case \cite{Golinelli2006}.

Our second example is a non-Hermitian version of Eq.~\eqref{hubbard}, that is, a modified Hubbard model of the form
\begin{equation}
\label{nhhubbard}
\begin{split}
\mathcal{H}_{\mathrm{NH}}=-&\sum_{k=1}^{\tilde{N}}t_k\left[e^{\alpha}c_{k+1}^\dagger c_k+e^{-\alpha}c_k^\dagger c_{k+1}\right]\\&+\sum_{k=1}^{\tilde{N}}V_k\left(c_{k+1}^\dagger c_{k+1}-\frac{1}{2}\right)\left(c_{k}^\dagger c_{k}-\frac{1}{2}\right),
\end{split}
\end{equation}
with $\alpha >0$. Although not quantum, non-Hermitian models of this type have attracted a lot of attention recently, both theoretically and experimentally, as they effectively describe quantum systems that are coupled to their environment \cite{Nakagawa2018,Zhao2019,Yang2019,Song2019,Mu2019}.

As one could suspect, those two models can both be described by the directed versions $X_{\rightleftarrows}(\mathfrak{S}_{(N-M,M)}\subset\mathfrak{S}_N, \tilde{S}_C)$ of the $X(\mathfrak{S}_{(N-M,M)}\subset\mathfrak{S}_N, \tilde{S}_C)$ graphs (see Fig.~\ref{asym}(b)). Obviously, the fact that their Laplacian matrices $\Delta^{\nu}_{\rightleftarrows}$ are no longer symmetric makes their spectral properties more difficult to study -- in particular, their spectra are not necessarily real. However, the identification of each vertex as a tabloid, and therefore the algebraic structure, is, for instance, still valid (see Section \ref{secxxx}). Moreover, the mapping between asymmetric exclusion processes and non-Hermitian quantum models \cite{Spohn1992} is transparent within our framework, since they are both described by the spectral properties of $\Delta^{\nu}_{\rightleftarrows}$. A more precise study of this matrix is left for future work.

\section{Conclusion}

In this work, we have shown that a wide range of classical and quantum 1D exchange models are described by the same theoretical object, namely the Laplacian matrix of a Schreier graph associated with the permutation group. This unifying description allows one to identify the algebraic structure of the problem in a natural way. As a consequence, one may access some peculiar eigenvalues much more easily, such as the spectral radius and the spectral gap, which are associated with the rate to stationarity in the classical stochastic case, and with the speed at which an adiabatic protocol may be performed in the context of adiabatic quantum computing. In particular, our result on the energy gap of inhomogeneous ferromagnetic Heisenberg spin chains may be tested in experiments, and has important potential applications for quantum technologies.

Furthermore, we stressed that the graphs we defined in this article are deeply related to the celebrated Bethe ansatz, which can be regarded as a powerful tool in order to compute their spectra in unweighted case. As an illustration, we obtained the spectral radii of Schreier graphs associated with the permutation group with a large number of elements. More generally, we believe these graphs have a central importance in the theory of quantum integrable systems. Nevertheless, we lay emphasis on the fact that our description goes beyond the integrable case, as it also enables one to describe completely inhomogeneous systems.

There are many open questions on these mathematical objects, which once tackled would give crucial information about all the models we described above. Moreover, an interesting and ambitious perspective for future works, besides the ones mentioned in this paper, would be to extend this framework to integer spin-chains. Then, the fact that our method is efficient in order to study the energy gap suggests us that it could be well adapted to tackle the celebrated Haldane conjecture on antiferromagnetic integer spin chains \cite{Haldane1983a,Haldane1983b}. 

In conclusion, this work stands at the crossroad between quantum physics, classical physics, and pure mathematics. This connection sheds a new light on some well-known problems, and paves the way to fruitful collaborations between different research fields.

\begin{acknowledgments}
This work has been partially funded by the Singapore NRF Grant numbers NRF-NRFI2017-04 (WBS No. R-144-000-378-281) and NRF-NRFF2018-02.
\end{acknowledgments}

\appendix

\section{Notions of graph theory}

\label{apbasicgraph}

In this section, we recall some basic definitions and properties of graph theory \cite{Bondy2008}. A graph 
$\mathcal{G}=(\mathcal{V},\mathcal{E})$ is defined by a set of vertices $\mathcal{V}$ and a set of edges $\mathcal{E}$ that are characterized by pair of vertices. When $(i,j)\in\mathcal{E}$ if and only if $(j,i)\in\mathcal{E}$, $\mathcal{G}$ is said to be undirected. A graph is weighted when each edge $e$ is associated with a real number $w_e$ such that all $w_e$ are not necessary equal, and unweighted when $w_e=1$ for all edges $e$. The degree $\mathrm{deg}(v)$ of a vertex $v$ is then the sum of the weights of all the edges between $v$ and its adjacent vertices. When all the vertices of a graph have the same degree $r$, the graph is said to be $r-$regular. A bipartite graph is a graph such that its vertices can be divided into two disjoint sets $\mathcal{V}_1$ and $\mathcal{V}_2$ where each edge connects an element of $\mathcal{V}_1$ with an element of $\mathcal{V}_2$. A graph $\mathcal{G}_1=(\mathcal{V}_1,\mathcal{E}_1)$ is a covering graph of $\mathcal{G}_2=(\mathcal{V}_2,\mathcal{E}_2)$ if we can map $\mathcal{G}_2$ into $\mathcal{G}_1$ while respecting the topology of the graph, i.e., if there is a surjective map $f:\mathcal{V}_1\to\mathcal{V}_2$ such that, for each $v\in\mathcal{V}_1$, the restriction of $f$ to a neighbourhood of $v$ is a bijection onto a neighbourhood of $f(v)$.

The main branch of graph theory is arguably the so-called spectral graph theory, which consists in associating matrices to the graphs of interest and then relating the properties of the graph to the spectra of their matrices \cite{Brouwer2012}. The simplest matrix that one can associate with a graph $\mathcal{G}$ is its adjacency matrix $A_{\mathcal{G}}$, which is written in the basis of the vertices such that entry $A_{ij}$ is  equal to $w_{(i,j)}$ if vertex ``$i$'' is adjacent to vertex ``$j$'' and $0$ otherwise. Another simple matrix that can be written in the same basis is the degree matrix $D_{\mathcal{G}}$, which is a diagonal matrix whose diagonal entries are given by the degree of the corresponding vertex. Then, the Laplacian matrix is defined by $\Delta_{\mathcal{G}}=D_{\mathcal{G}}-A_{\mathcal{G}}$. For instance, one can check that the Laplacian matrices of the graphs in panels (a) and (b) of Fig.~\ref{ex4} of the main text are respectively given by:
\begin{equation*}
\left(\begin{smallmatrix}
			J_2 & -J_2 & 0 & 0 & 0 & 0 \\
			-J_2 & J_1+J_2+J_3 & -J_3 & -J_1 & 0 & 0\\
			0 & -J_3 & J_1+J_3 & 0 & -J_1 & 0\\
			0 & -J_1 & 0 & J_1+J_3 & -J_3 & 0\\
			0 & 0 & -J_1 & -J_3 & J_1+J_2+J_3 & -J_2\\
			0 & 0 & 0 & 0 & -J_2 & J_2
			\end{smallmatrix}\right),
\end{equation*}
and
\begin{equation*}
\left(\begin{smallmatrix}
		J_2+J_4 & -J_2 & 0 & 0 & -J_4 & 0 \\
		-J_2 & J_1+J_2+J_3 +J_4& -J_3 & -J_1 & 0 & -J_4\\
		0 & -J_3 & J_1+J_3 & 0 & -J_1 & 0\\
		0 & -J_1 & 0 & J_1+J_3 & -J_3 & 0\\
		-J_4 & 0 & -J_1 & -J_3 & J_1+J_2+J_3+J_4 & -J_2\\
		0 & -J_4 & 0 & 0 & -J_2 & J_2+J_4
		\end{smallmatrix}\right).
\end{equation*}
Up to a minus sign, $\Delta_{\mathcal{G}}$ can be seen as a discrete version of the continuous Laplacian. Moreover, it is closely linked to the stochastic properties of the graph, as it is can be related to the transition matrix of a random walk on $\mathcal{V}$ \cite{Lovász93randomwalks}.

In the case of an undirected graph, $\mathcal{G}$, $\Delta_{\mathcal{G}}$ is symmetric: Thus, it can be diagonalized in an orthogonal basis and its  spectrum $\sigma\left(\Delta_{\mathcal{G}}\right)$ is real. From the positivity of the diagonal entries of $\Delta_{\mathcal{G}}$ and the fact that each of them is equal to the sum of the absolute values of the non-diagonal entries in their respective row, one can easily deduce that $\sigma\left(\Delta_{\mathcal{G}}\right)$ is also positive. Since every row sum is equal to zero,
the vector $u=(1,1,\ldots,1)$ always satisfies $\Delta_{\mathcal{G}}u=0$, we see that   $\sigma\left(\Delta_{\mathcal{G}}\right)$ always contains $0$. In fact, it is easy to show that the multiplicity of $0$ is equal to the number of connected components of $\mathcal{G}$. Therefore, we see that the second smallest eigenvalue $\lambda_*$, or spectral gap, is closely related to the topology of $\mathcal{G}$, which explains why this quantity has been intensively studied by mathematicians \cite{Chung1996,Godsil2001}. Intuitively, $\lambda_*$ can be seen as a measure of the connectivity of $\mathcal{G}$. Another quantity related to the geometrical properties of the graph is the spectral radius $\rho\left(\Delta_{\mathcal{G}}\right)$, which in our case is equal to the largest eigenvalue of $\Delta_{\mathcal{G}}$. One important property in the case of a connected graph is e.g., \cite{Mohar1991}
\begin{equation}
\label{biregmax}
\rho\left(\Delta_{\mathcal{G}}\right)\le\max\{\mathrm{deg}(u)+\mathrm{deg}(v);(u,v)\in\mathcal{E}\},
\end{equation}
with equality if $\mathcal{G}$ is bipartite and regular.

\section{Representations of $\mathfrak{S}_N$}
\label{aprep}

A complete description of the set $\mathrm{IR}$ of irreducible representation (irreps) of the permutation group $\mathfrak{S}_N$ can be found in Ref. \cite{James1984}. 

The set $\mathrm{IR}$ is in bijection with the conjugacy classes of $\mathfrak{S}_N$ (2 elements $g_1$ and $g_2$ of a group are conjugate if there is an element $h$ in the group such that $g_1 = h^{-1} g_2h$), which are uniquely labeled by the different structures of their decomposition in disjoint cyclic permutations. Therefore, there is a one-to-one correspondence between $\mathrm{IR}$ and the set of partitions of $N$. Equivalently, each partition $\mu=\left[\mu_1,\ldots,\mu_r\right]$  of $N$ can be graphically represented by a Young diagram, which is a left-justified set of boxes with $r$ rows such that each row contains $\mu_i$ boxes. For example, one of the irreps of $\mathfrak{S}_9$ is characterized by the partition $\left[5,3,1\right]$, or equivalently by the following Young diagram:
\begin{equation}
Y_{\left[5,3,1\right]}\equiv\yng(5,3,1).
\end{equation}
$\left[5,3,1\right]$ is also called the shape of $Y_{\left[5,3,1\right]}$. The conjugate of a Young diagram of shape $\mu=\left[\mu_1,\ldots,\mu_r\right]$ is the diagram with columns of lengths $\mu_1,\ldots,\mu_r$. For example, the conjugate of the previous example is
\begin{equation}
Y_{\left[3,2,2,1,1\right]}\equiv\yng(3,2,2,1,1).
\end{equation}

The so-called dominance order $\trianglerighteq$ on Young diagrams is such that $\mu\equiv\left[\mu_1,\ldots,\mu_r\right]\trianglerighteq\nu\equiv\left[\nu_1,\ldots,\nu_r\right]$ (where the last terms of one of the partitions may be equal to zero) if
\begin{equation}
\mu_1+\cdots+\mu_k\ge\nu_1+\cdots+\nu_k\quad\text{for all }1\le k\le r.
\end{equation}
For example, one has
\begin{equation}
\yng(4,3,1,1)\quad\trianglerighteq\quad\yng(3,2,1,1,1,1)~.
\end{equation}
In more visual terms, it means that one can go from the left diagram to the right one by moving a certain number of boxes from upper rows to lower rows and right to left. Note that the dominance order is only a partial order when $N>5$. For instance, it is not possible to compare
\begin{equation}
\yng(2,2,2)\quad\text{and}\quad\yng(3,1,1,1)~.
\end{equation}

Given a Young diagram, one may associate different Young tableaux by labeling its boxes by integers. A tableau is said to be standard when its entries are increasing from left to right along the rows and up to down along the columns.

Two tableaux of same shape $\mu$ are said to be row-equivalent if they are equal up to permutations of integers belonging to the same rows. For example,
\begin{equation}
\young(84519,372,6)\quad\text{and}\quad\young(19854,732,6)
\end{equation}
are row-equivalent. Then, we define tabloids as the row-equivalence classes among the set of $\mu$-tableaux. For the previous example, the corresponding tabloid may be graphically represented as
\ytableausetup{tabloids,centertableaux}
\begin{equation}
\label{extabloid}
\ytableaushort{14589,237,6}~.
\end{equation}
Note than the number of $\mu$-tabloids is $D_{\mu}$, as defined in Eq.~\eqref{dnu} of the main text. The permutation module is then defined as the $\mathbb{C}$-vector space $M^{\mu}$ whose basis is indexed by the set of $\mu$-tabloids. This vector space is crucial for the construction of $\mathrm{IR}$. As an important peculiar case, $M^{\left[1,1,\ldots,1\right]}$ is equivalent to $\mathbb{C}\left[\mathfrak{S}_N\right]$, the group algebra of $\mathfrak{S}_N$, whose basis $(e_P)_{P\in \mathfrak{S}_N}$ is indexed by the elements of $\mathfrak{S}_N$. Additionally, the internal multiplication on $\mathbb{C}\left[\mathfrak{S}_N\right]$ verifies $e_P\cdot e_Q =e_{PQ}$. Then, one can naturally associate the regular representation $\mathcal{R}$ with $\mathbb{C}[\mathfrak{S}_N]$ by considering the canonical action of $\mathfrak{S}_N$ on $\mathbb{C}[\mathfrak{S}_N]$: Writing a vector in $\mathbb{C}[\mathfrak{S}_N]$ as $u=\sum_{Q\in \mathfrak{S}_N}u_Qe_Q$ with $u_Q\in \mathbb{C}$, and given $P\in \mathfrak{S}_N$, the linear map $\mathcal{R}(P)$ is defined by
\begin{equation}
\mathcal{R}(P)\cdot u=e_{P}u=\sum_{Q\in \mathfrak{S}_N}u_Qe_{PQ}=\sum_{Q\in \mathfrak{S}_N}u_{P^{-1}Q}e_{Q},
\end{equation}
which can be linearly extended on a representation of $\mathbb{C}[\mathfrak{S}_N]$ by writing $\mathcal{R}\left(\sum_{Q\in \mathfrak{S}_N}u_Qe_Q\right)\equiv\sum_{Q\in \mathfrak{S}_N}u_Q\mathcal{R}(Q)$. Similarly, one can identify any permutation module $M^{\mu}$ with a representation of $\mathfrak{S}_N$ by considering the natural action of $\mathfrak{S}_N$ on the vector space $M^{\mu}$.

Given a $\mu$-tableau $T$, one may associate the following element of $M^{\mu}$ known as a polytabloid:
\begin{equation}
\label{etspecht}
E_T=\sum_{P\in C_T}\epsilon(P)\{P(T)\},
\end{equation}
where $C_T$ is the set of permutations preserving all columns of $T$, $\epsilon(P)$ is the sign of the permutation $P$, and $\{T\}$ is the tabloid corresponding to $T$. The Specht module $S_\mu$ is then defined as the subspace of $M_{\mu}$ generated by all the elements $E_T$ when $T$ runs through all the $\mu$-tableaux. It can be shown that the basis of $S^\mu$ is given by the elements $E_T$ when $T$ runs though all the standard Young tableaux of shape $\mu$.  In particular, its dimension is given by the number of standard Young tableaux, which is given by the so-called hook length formula. Furthermore, a crucial theorem known as Young's rule states that the permutation module $M^\mu$ can be decomposed in the following way:
\begin{equation}
\label{youngsruleapp}
M^\mu\cong\bigoplus_{\mu'\trianglerighteq\mu}k_{\mu'\mu}S^{\mu'}\quad\quad (k_{\mu\mu}=1),
\end{equation}
where $k_{\mu'\mu}$ are positive integers called the Kostka numbers.

As it is the case for the permutation modules and the group algebra, $S^\mu$ can be identified with a representation of $\mathfrak{S}_N$ by considering the natural action of $\mathfrak{S}_N$ on $S^\mu$. Then, it can be shown that the set of all the Specht modules $S^\mu$ is, in fact, $\mathrm{IR}$. Note that when taking  $\mu=\left[1,1,\ldots,1\right]$, Eq.~\eqref{youngsruleapp} implies in particular that the regular representation is a sum of all the irreps. In this case, the $k_{\mu'\left[1,1,\ldots,1\right]}$ numbers are given by the dimensions of the $S^{\mu'}$ irreps. Intuitively, keeping in mind Eq.~\eqref{etspecht}, an irrep $S_{\mu}$ can be seen as symmetrizing the rows and anti-symmetrizing the columns of the $\mu$-tableaux \cite{notenotationirrep}.


\begin{thebibliography}{100}
\bibitem{Diaconis} P. Diaconis, {\em Group Representations in Probability and Statistics} (Institute of Mathematical Statistics, Hayward, 1988).
\bibitem{Volosniev2014} A.G. Volosniev, D.V. Fedorov, A.S. Jensen, N.T. Zinner, and M. Valiente, Strongly interacting confined quantum systems
in one dimension, {\em Nat. Commun.} {\bf 5}, 5300 (2014).
\bibitem{Deuretzbacher2014} F. Deuretzbacher, D. Becker, J. Bjerlin, S. M. Reimann, and L. Santos, Quantum magnetism without lattices in strongly interacting one-dimensional spinor gases, {\em Phys. Rev. A} {\bf 90}, 013611 (2014).
\bibitem{Heisenberg1928} W. Heisenberg, Zur Theorie des Ferromagnetismus, {\em Z. Physik} \textbf{49}, 619 (1928).
\bibitem{MacDonald1969} C. T. MacDonald, and J. H. Gibbs, Concerning the kinetics of polypeptide synthesis on polyribosomes, {\em Biopolymers} \textbf{7}, 707 (1969).
\bibitem{Mattis1981} D. C. Mattis, {\em The Theory of Magnetism, vol. 1} (Springer-Verlag, Berlin, 1981).
\bibitem{Bethe1931} H. Bethe, Zur Theorie der Metalle. I. Eigenwerte und Eigenfunktionen der linearen Atomkette, {\em Z. Physik} {\bf 71}, 205 (1931).	
\bibitem{Lieb1963} E. H. Lieb, and W. Liniger, Exact Analysis of an Interacting Bose Gas. I. The General Solution and the Ground State, {\em Phys. Rev.} {\bf 130}, 1605 (1963).
\bibitem{Yang1967} C. N. Yang, Some Exact Results for the Many-Body Problem in one Dimension with Repulsive Delta-Function Interaction, {\em Phys. Rev. Lett.} {\bf 19}, 1312 (1967).
\bibitem{Sutherland1968} B. Sutherland, Further Results for the Many-Body Problem in One Dimension, {\em Phys. Rev. Lett.} {\bf 20}, 98 (1968).
\bibitem{Andrei1983} N. Andrei, K. Furuya, and J.H. Lowenstein, Solution of the Kondo problem, {\em Rev. Mod. Phys.} \textbf{55}, 331 (1983).
\bibitem{Ambjorn2006} J. Ambj{\o}rn, R.A. Janik, and C. Kristjansen, Wrapping interactions and a new source of corrections to the spin-chain/string duality, {\em Nucl. Phys. B} \textbf{736}, 288 (2006).
\bibitem{Baxter1971} R. J. Baxter, Eight-Vertex Model in Lattice Statistics, {\em Phys. Rev. Lett.} {\bf 26}, 832 (1971).
\bibitem{Batchelor1995} M. T. Batchelor, R. J. Baxter, M. J. O'Rourke, and C. M. Yung, Exact solution and interfacial tension of the six-vertex model with anti-periodic boundary conditions, {\em J. Phys. A} \textbf{28}, 2759 (1995).
\bibitem{Golinelli2006} O. Golinelli, and K. Mallik, The asymmetric simple exclusion process: an integrable model for non-equilibrium statistical mechanics,  {\em J. Phys. A: Math. Gen.} \textbf{39}, 12679 (2006).
\bibitem{Sutherland2004} B. Sutherland, {\em Beautiful models: 70 years of exactly solved quantum many-body problems} (World Scientific Press, Singapore, 2004).
\bibitem{Korepin1993} V. E. Korepin, N. M. Bogoliubov, and A. G. Izergin, {\em Quantum Inverse Scattering Method and Correlation Functions} (Cambridge University Press, Cambridge, 1993).
\bibitem{Sutherland1975} B. Sutherland, Model for a multicomponent quantum system, {\em Phys. Rev. B} {\bf 12}, 3795 (1975).
\bibitem{LiebMattisPR} E. Lieb and D. Mattis, Theory of Ferromagnetism and the Ordering of Electronic Energy Levels, {\em Phys. Rev.} {\bf 125}, 164 (1962).
\bibitem{Albash2018} T. Albash and D. A. Lidar, Adiabatic quantum computation, {\em Rev. Mod. Phys.} {\bf 90}, 015002 (2018).
\bibitem{Decamp2020} J. Decamp, J. Gong, H. Loh, and C. Miniatura, Graph-theory treatment of one-dimensional strongly repulsive fermions, {\em Phys. Rev. Res.} {\bf 2}, 023059 (2020).
\bibitem{Dirac1958} P. A. M. Dirac, {\em The Principles of Quantum Mechanics, 4th Ed.} (Clarendon Press, Oxford, 1958).
\bibitem{Pfeifer2003} W. Pfeifer, {\em The Lie Algebras $SU(N)$: An Introduction} (Birkh\"{a}user Basel, 2003).
\bibitem{Brouwer2012} A. E. Brouwer and W. H. Haemers, {\em Spectra of Graphs} (Springer, New York, 2012).
\bibitem{notegenerating} A subset $S$ of a multiplicative group $G$ is generating if one can obtain any element of $G$ by multiplying elements of $S$.
\bibitem{Fulton2004} W. Fulton and J. Harris, {\em Representation Theory} (Springer, New York, 2004).
\bibitem{Liebeck} G. James and M. Liebeck, {\em Representations and Characters of Groups (2nd ed.)} (Cambridge University Press, Cambridge, London, 2001).
\bibitem{Hamermesh_book} M. Hammermesh, {\em Group theory and its applications to physical problems} (Dover, New York, 1989).
\bibitem{Noteirrep} For the construction of the irreducible representations of the permutation group $\mathfrak{S}_N$ and the definitions of the associated concepts (tabloid, permutation module, Young's rule), see Appendix \ref{aprep}.
\bibitem{James1984} G. James and A. Kerber, {\em The Representation Theory of the Symmetric Group} (Cambridge University Press, 1984).
\bibitem{Poignard2018} C. Poignard, T. Perreira, and J, P. Pade, Spectra of Laplacian Matrices of Weighted Graphs: Structural Genericity Properties, {\em SIAM J. Appl. Math.} {\bf 78}, 372 (2018).
\bibitem{Bacher1994} R. Bacher, Valeur propre minimale du laplacien de Coxeter pour le groupe sym\'{e}trique, {\em J. Albegra} {\bf 167}, 460 (1994).
\bibitem{Anderson1985} W. N. Anderson, and T. D. Morley, Eigenvalues of the Laplacian of a graph, {\em Linear Multilinear A} {\bf 18}, 141 (1985).
\bibitem{Jordan1928} P. Jordan, and E. Wigner, \"{U}ber das Paulische \"{A}quivalensverbot, {\em Z. Physik} {\bf 47}, 631 (1928). 
\bibitem{Lovász93randomwalks} L. Lov\'{a}sk, {\em Random walks on graphs: A survey} (J\'{a}nos Bolyai Math. Soc., Budapest, 1996).
\bibitem{Diaconis1993} P. Diaconis and L. Saloff-Coste, Comparison theorems for reversible Markov chains, {\em Ann. Appl. Prob.} \textbf{3}, 696 (1993).
\bibitem{Ng1996} L. L. Ng, {\em Heisenberg Model, Bethe Ansatz, and Random Walks} (Senior Honor Thesis, Harvard University, 1996).
\bibitem{notejw} Notice that the classical equivalence between interchange and exclusion processes is analogous to the mapping between quantum Heisenberg spin-chains and the Hubbard model via the Jordan-Wigner transformation.
\bibitem{Richards1977} P. M. Richards, Theory of one-dimensional hopping conductivity and diffusion, {\em Phys. Rev. B} \textbf{16}, 1393 (1977).
\bibitem{Krug1997} J. Krug, Origins of scale invariance in growth processes, {\em Adv. Phys.} \textbf{46}, 139 (1997).
\bibitem{Wolf1998} M. Schreckenberg, and D. E. Wolf, {\em  Traffic and granular flow ’97} (Springer-Verlag, New-York, 1998).
\bibitem{Klumpp2003} S. Klumpp, and R. Lipowsky, Traffic of molecular motors through tube-like compartments, {\em J. Stat. Phys.} \textbf{113}, 233 (2003).
\bibitem{Nakagawa2018} M. Nakagawa, N. Kawakami, and M. Ueda, Non-hermitian kondo effect in ultracold alkaline-earth atoms, {\em Phys. Rev. Lett.} \textbf{121}, 203001 (2018).
\bibitem{Zhao2019} H. Zhao, X. Qiao, T. Wu, B. Midya, S. Longhi, and L. Feng, Non-hermitian topological light steering, {\em Science} \textbf{365}, 1163 (2019).
\bibitem{Yang2019} Z. Yang and J. Hu, Non-hermitian hopf-link exceptional line semimetals, {\em Phys. Rev. B} \textbf{99}, 081102 (2019).
\bibitem{Song2019} F. Song, S. Yao, and Z. Wang, Non-hermitian skin effect and chiral damping in open quantum systems, {\em Phys.
Rev. Lett.} \textbf{123}, 170401 (2019).
\bibitem{Mu2019} S. Mu, C. H. Lee, L. Li, and J. Gong, Emergent Fermi surface in a many-body non-Hermitian Fermionic chain, arXiv:1911.00023 (2019).
\bibitem{Spohn1992} L.-H. Gwa, and H. Spohn, Bethe solution for the dynamical exponent of the noisy Burger equation, {\em Phys. Rev. A} \textbf{92}, 844 (1992).
\bibitem{Haldane1983a} F. D. M. Haldane, Nonlinear Field Theory of Large-Spin Heisenberg Antiferromagnets: Semiclassically Quantized Solitons of the One-Dimensional Easy-Axis N\'eel State, {\em Phys. Rev. Lett.} {\bf 50}, 1153 (1983).
\bibitem{Haldane1983b} F. D. M. Haldane, Continuum dynamics of the 1-D Heisenberg antiferromagnet: Identification with the O(3) nonlinear sigma model, {\em Phys. Lett. A} {\bf 93}, 464 (1983).


\bibitem{Bondy2008} J. Bondy and U. Murty, {\em Graph Theory} (Springer-Verlag, London, 2008).
\bibitem{Chung1996} F. Chung, {\em Spectral Graph Theory} (American Mathematical Society, 1996).
\bibitem{Godsil2001} C. Godsil and G. Royle, {\em Algebric Graph Theory} (Springer, New York, 2001).
\bibitem{Mohar1991} B. Mohar, {\em The Laplacian spectrum of graphs}  (Wiley, 1991).
\bibitem{notenotationirrep} As it is typically done, we use (quite abusively) the same notation for the permutation/Specht modules and the corresponding representations.
\end{thebibliography}
\end{document}